\documentclass{article}
\usepackage{lineno}
\usepackage{xcolor}
\usepackage{amsmath}

\usepackage{PRIMEarxiv}

\usepackage[utf8]{inputenc} 
\usepackage[T1]{fontenc}    
\usepackage{hyperref}       
\usepackage{url}            
\usepackage{booktabs}       
\usepackage{amsfonts}       
\usepackage{nicefrac}       
\usepackage{microtype}      
\usepackage{lipsum}
\usepackage{fancyhdr}       
\usepackage{graphicx}       
\graphicspath{{media/}}     

\title{Measurement of Time Resolution of Scintillation Detectors with EQR-15 Silicon Photodetectors for the Time-of-Flight Neutron Detector of the BM@N Experiment}

\author{
  F. Guber, A. Ivashkin, N. Karpushkin, A. Makhnev, S. Morozov, D. Serebryakov \\
  Institute for Nuclear Research of the Russian Academy of Sciences \\
  60-letiya Oktyabrya prospekt 7a, Moscow 117312, Russia\\
  \texttt{guber@inr.ru} \\
  \AND
  V. Baskov, V. Polyansky \\
  P.N. Lebedev Physical Institute of the Russian Academy of Sciences \\
  Leninskiy Prospekt 53, Moscow 119991, Russia \\
}

\begin{document}
\maketitle

\begin{abstract}
  To study the dependence of the equation of state of high density nuclear matter on the term characterizing the isospin (proton-neutron) asymmetry of nuclear matter, it is necessary to measure azimuthal flow of neutrons as well as azimuthal flow of charged particles from a dense nuclear matter in the nuclear-nuclear collisions. For this purpose INR RAS is developing a new high-granular neutron detector which will be used in the BM@N experiment at the extracted beam of the Nuclotron accelerator at JINR (Dubna). This detector will identify neutrons and measure their energies in the heavy-ion collisions up to 4 GeV per nucleon.

  This article presents the results of measurements of the time resolution and light yields of samples of scintillation detectors with sizes 40$\times$40$\times$25 mm$^3$ that will be used in a neutron detector based on the currently available fast plastic scintillator manufactured by JINR using an EQR15 11-6060D-S photodetector for light readout. For comparison, the results of measurements for a detector of the same size with a fast scintillator EJ-230 and with the same type of photodetector are given. The measurements were made on cosmic muons as well as on the electron synchrotron ''Pakhra'' of the Lebedev Physical Institute of the Russian Academy of Sciences located in Troitsk, Moscow.
\end{abstract}



\section{Introduction}
\label{sec:intro}
Research program of the BM@N experiment (Baryonic Matter at Nuclotron) at JINR (Dubna)   includes studying the equation of state of high density nuclear matter. This matter forms during collisions of heavy ions with a fixed target at incident ion energies of up to 4 AGeV \cite{Kapishin:2020cwk, Senger:2022bzm}. In addition to measuring the azimuthal asymmetry of charged particle flows, understanding the equation of state's dependence on isospin (proton-neutron) asymmetry \cite{Sorensen:2023zkk} requires measuring neutron flows during nucleus-nucleus collisions. A previous experiment with gold ion collisions at 400 AMeV conducted more than a decade ago at GSI \cite{Russotto:2011hq} revealed the equation of state's sensitivity to the direct and elliptical neutron flows in relation to corresponding flows of charged hadrons.

The BM@N experiment provides a unique opportunity to measure the neutron and proton flow ratio at nuclear collision energies, where nuclear density is 2-4 times higher than usual. This density is comparable to that generated during neutron star mergers. Whilst the BM@N magnetic spectrometer measures proton flows, a new high granular time-of-flight neutron detector, HGN (High Granularity Neutron Detector), is under development at INR RAS for identifying and measuring energies of neutrons \cite{Guber_arxiv}.

The HGN detector can be placed within the BM@N experimental hall at a maximum distance of about 5m from the target. Given the relatively short time-of-flight base, the active elements of the detector should have a time resolution of approximately 100-150 ps. This precision is necessary to achieve an acceptable energy resolution for detected neutrons with kinetic energies up to 4 GeV. To meet these requirements, plastic scintillators with dimensions of 40$\times$40$\times$25 mm$^3$ were chosen as the active components for the HGN detector. The light from scintillators is detected by a single silicon photomultiplier with an active area of 6$\times$6 mm$^2$. The neutron detector consists of about 2000 such scintillation detectors. During the selection of scintillator material and photodetector type, both the time resolution demands of scintillation detectors and cost optimization must be taken into account. In this regard, preliminary test measurements have been done using cosmic muons, and results for various scintillation detector samples assembled with scintillators and silicon photodetectors from different manufacturers have been published \cite{Guber:2023qjk}.

This article presents the measurements of the time resolution and light yields of scintillation detector samples, based on the currently most accessible fast plastic scintillator produced at JINR, in combination with the presently available photodetector model EQR15 11-6060D-S \cite{NDL-EQR15}. The measurements were performed with cosmic muons as well as with the electron beam of the ''Pakhra'' synchrotron (LPI, Troitsk, Moscow) \cite{Alekseev:2019sxi}. The measurement results are compared with those obtained for detector samples featuring the EJ-230 fast scintillator \cite{EJ-230} with the same photodetector.

\section{Measurement of time resolution of scintillation detectors with cosmic muons}
For measuring the scintillation detector time resolution, several samples were manufactured with dimensions of 40$\times$40$\times$25 mm$^3$. These samples are polystyrene-based scintillators with 1.5\% paraterphenyl and 0.01\% POPOP. The decay time of this scintillator was measured at 3.9$\pm$0.7 ns. For comparison, the time resolution was also determined for detector samples based on the faster EJ-230 scintillator, featuring a decay time of 2.8$\pm$0.5 ns and identical dimensions. A silicon photodetector with a sensitive area of 6$\times$6 mm$^2$ was positioned at the center of a larger face of  scintillator. The remaining area of this face was covered with a light-absorbing black tape. The other faces were coated with a white diffuse reflector of titanium oxide ($TiO_2$). The photodetector is mounted on PCB (printed circuit board) with the preamplifier and the discriminator. Both the scintillator and the PCB are placed in light-protective plastic cases produced by 3D printing.

Modern fast silicon photodetectors EQR15 11-6060D-S \cite{NDL-EQR15}, recently introduced to the market and showing a quantum efficiency of 45\% along with a gain of approximately $4\times10^5$, were used as the photodetectors. The signal from the scintillation detector is amplified with a preamplifier featuring an LMH6629MF op-amp (with a gain of 20 dB and a bandwidth of 600 MHz at a 3 dB level, and noise of <2.2 $nV/\sqrt{Hz}$). To measure the response time, the signal from the preamplifier was connected to a rapid discriminator (ADCMP553) with a fixed threshold. The output of this discriminator was then digitized for subsequent analysis using a precise ADC CAEN DT5742, offering a sampling step of 200 ps. The inherent time resolution of the photodetector with an amplifier was measured using a picosecond laser. The resulting spread in the delay between the laser trigger signal and the photodetector response measured 30–40 psec \cite{Guber:2023qjk}. A microchannel PMT, with a 20 mm thick quartz radiator (Photonis XP85012-FIT/Q) \cite{Karavicheva_2017}, functioning as a Cherenkov detector, was used as a starting (trigger) detector. The time resolution of this detector is about 22 psec. The signal from the start detector was also digitized using a CAEN DT5742 ADC. Time resolution measurements on cosmic muons for the scintillation detector with a JINR plastic scintillator and an EQR15 11-6060D-S photodetector shows a value of 125 ps (see Figure \ref{fig:resol} on the left), whereas the time resolution for a scintillation detector with EJ-230 scintillator and the same photodetector is 105 psec.

\section{Measurement of time resolution of scintillation detectors using the electron beam of the ''Pakhra'' synchrotron (LPI, Troitsk)}
\begin{figure}[htbp]
  \centering
  \includegraphics[width=.8\textwidth]{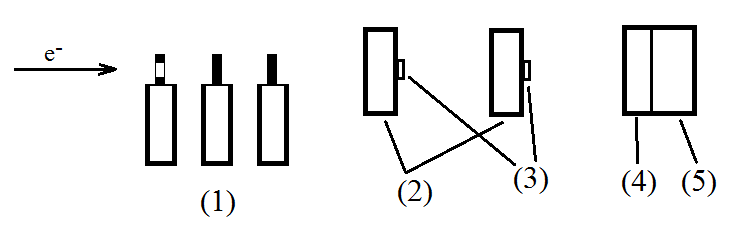}
  \caption{Schematic view of the setup for measurements of detector characteristics using the electron beam of the ''Pakhra'' synchrotron (LPI, Troitsk). (1) Beam counters, (2) Scintillator samples, (3) Photodiodes, (4) Quartz Cherenkov radiator, (5) MCP PMT.}
  \label{fig:setup}
\end{figure}

To measure the time resolution of scintillation detector samples using an electron beam, the experimental setup has been assembled (see Figure \ref{fig:setup}). The electron beam with an energy of 280 MeV, was directed into the experimental area. A veto counter, followed by two scintillation beam detectors with transverse dimension of 10$\times$10 mm$^2$ and a thickness of 5 mm (1), were placed on the beam axis and used in the trigger. Subsequently, the scintillation detectors under study (2), along with the installed photodetectors (3), were positioned further along the beam, accompanied by a fast trigger detector consisting of a quartz radiator (4) and a microchannel PMT (5). 

\begin{figure}[htbp]
  \centering
  \includegraphics[width=.45\textwidth]{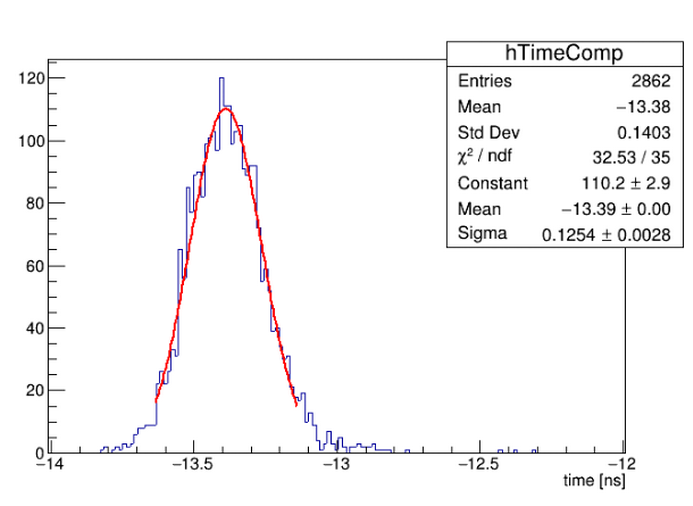}
  \qquad
  \includegraphics[width=.45\textwidth]{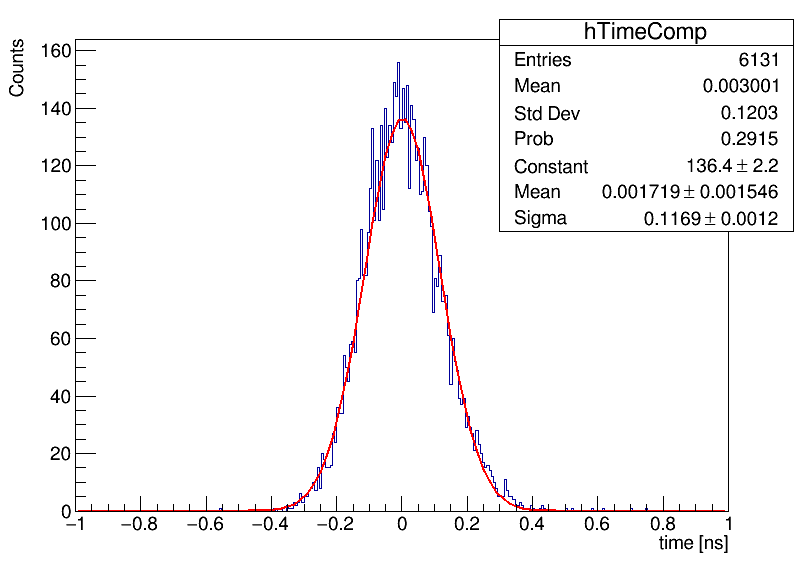}
  \caption{The results of time resolution measurements performed with cosmic muons (left) and an electron beam (right) for a scintillation detector equipped with a JINR scintillator and an EQR15 11-6060D-S photodetector.}
  \label{fig:resol}
\end{figure}

Time resolution measurement for a scintillation detector using an electron beam revealed a value of 117 ps when employing a JINR plastic scintillator with the EQR15 11-6060D-S photodetector (Figure \ref{fig:resol}, right). In comparison, the time resolution for a scintillation detector with EJ-230 scintillator and the same photodetector shows a value of 74 ps. The spectra show the time difference between comparator readings of the photodetector signal and the signal from the fast trigger Cherenkov counter (MCP). Furthermore, the light yield of scintillation detectors, measured in photoelectrons, was assessed for minimally ionizing particles (MIP) using the method outlined in \cite{Guber:2023qjk}. Namely, the parameter of the relationship between signal spread and its amplitude for the EQR15 11-6060D-S photodetector yielded a conversion factor of 1.8 mV/photoelectron. This was in contrast to earlier measurements with Hamamatsu and SensL photodetectors, which reported values of 0.47 mV/p.e. and 0.68 mV/p.e., respectively \cite{Guber:2023qjk}. It's important to highlight that the MIP signal's amplitude peak yielded a light output value of 158$\pm$9 photoelectrons for the detector equipped with the JINR scintillator, and 292$\pm$2 photoelectrons for the detector with the EJ-230 scintillator. The light output for the EJ-230 scintillator was nearly twice as high, leading to the enhanced time resolution of the detector using this scintillator. Additionally, the influence of signal shape on the conversion factor was considered. Measurements of signal shape for the JINR scintillator indicated a signal area (charge) to amplitude ratio of 18.22, whereas for the EJ-230 scintillator, this value was 18.01. Consequently, the measurement inaccuracy for the conversion factor reached approximately 1.1\%, comparable to the error in measuring MIP signal light output.

Scans were also conducted using an electron beam across the surface of the scintillators. Testing various points on the surface of the JINR scintillator revealed that the ratio of light output for particles passing through the center of the scintillator to those passing along its edge is 1.45. However, the time resolution at the edge deteriorates by 6\%. This discrepancy explains the slightly worse time resolution obtained from cosmic muon measurements compared to measurements conducted with an electron beam. In the case of cosmic muons, the tracks traverse the entire surface of the scintillator, making the time resolution value an average across all points on the surface.

\section{Conclusion}
At the ''Pakhra'' synchrotron (LPI, Troitsk, Moscow), the time resolution of scintillation detector samples using an electron beam with an energy of 280 MeV was measured. Two types of plastic scintillators are tested: a JIRN (Dubna) produced scintillator and an EJ-230 scintillator. Both types were coupled with photodetectors EQR15 11-6060D-S for light readout. Using the JINR scintillator resulted in a time resolution of 120 ps, meeting the requirements for the HGN neutron detector under development. The EJ-230 scintillator, owing to its faster emission time, yielded a superior time resolution of 75 ps for the scintillation detector based on it. This advantage was observed under identical conditions during measurements on the electron beam. However, considering the higher cost associated with the EJ-230 scintillator, we propose to use JINR scintillators as the primary choice for the HGN detector. A time resolution of 100-150 ps is sufficiently acceptable for studying neutron flow anisotropy within the BM@N experiment.

\section*{Acknowledgments}
This work was conducted at INR RAS and supported by the Russian Science Foundation grant \textnumero 22-12-00132.

\bibliographystyle{elsarticle-num} 
\bibliography{references.bib}

\end{document}